\documentstyle[prl,aps,twocolumn]{revtex}
\begin{document}
\title{ Magnetism and superconductivity in underscreened Kondo chains}
\draft
\author{N. Andrei${}^\dagger$ and E. Orignac${}^{\dagger,*}$}
\address{${}^\dagger$ Serin Physics Laboratory, Rutgers University,
P. O. Box 849, Piscataway NJ 08855-0849 USA \\
${}^*$ Laboratoire de Physique Th\'eorique de l'Ecole Normale
Sup\'erieure, 24 Rue Lhomond, 75231 Paris Cedex 05}
\wideabs{
\date{\today}
\maketitle
\begin{abstract}
We present a one dimensional model of electrons coupled to localized 
moments of spin $S\ge 1$ in which
 magnetism and superconductivity interplay in a 
nontrivial manner. This model has a non-Fermi liquid ground state of
the chiral spin liquid type. A non-conventional
odd-frequency pairing is shown to be the dominant instability of the
system, together with antiferromagnetism of the local moments. 
We argue that this model captures  the physics of the
Kondo-Heisenberg spin $S=1$ chain, in the limit of strong Kondo
coupling. Finally, we discuss briefly the effect of interchain
coupling. 
\end{abstract}
}
The interplay of magnetism and superconductivity is a fundamental
problem in condensed matter physics. Superconductivity is associated
with the pairing of electron states  related by time
reversal. Magnetic states, in which time reversal symmetry is
lost, should therefore strongly compete with superconductivity. 
Thus it came as a surprise
when it was discovered in 1984 by Schlabitz et al.\cite{schlabitz86}
that magnetism and superconductivity actually coexisted in the heavy
fermion compound $\mathrm{URu_2Si_2}$. Since then, other heavy
fermion superconductors were shown to present magnetic moments in
their superconducting phase\cite{amato_heavy_musr}.  All these compounds
contain rare earth or actinide ions with very localized $4f$ or $5f$
orbitals, strongly interacting
 with the conduction band. (This is in contrast to such cases
as  the Chevrel phases where
magnetism and superconductivity coexist because the magnetic moments
responsible for magnetism are only very weakly coupled with the
electrons that form the condensate\cite{fischer_chevrel_phases}.)

The physics of heavy fermion
compounds  is believed to be described by the Kondo
Lattice model in which conduction electrons interact with the local
moments associated with  the localized $f$ electrons, and 
a large amount of theoretical work was carried out. However,
although the single impurity Kondo problem is well
understood\cite{kondo_papers}, the
theoretical analysis of the Kondo Lattice model has proven extremely
difficult.  This is because the Kondo effect (the
 quenching of the local moments)   competes, in the lattice
problem, with the
Rudermann-Kittel-Kasuya-Yosida interaction which orders the local
moments. This frustration is believed to be at the origin of the rich
phase diagram of heavy fermions systems. There exist at present
various theories of the superconducting ground state of the Kondo
Lattice Model. Conventional scenarios involve the pairing of fermionic
quasiparticles into either a spin triplet ``p-wave'' state or a spin
singlet ``d-wave'' state\cite{varma_vf_review}. A  less
conventional pairing\cite{coleman_three_body} involves the formation of 
a
 spin 
singlet, isotropic, odd in time  superconducting order parameter, known 
as
odd-frequency pairing\cite{berizinskii_odd_freq,balatsky_odd_freq}.

In one dimension, the theoretical situation is more
favorable\cite{tsunetsugu_kondo_1d}, 
since there are
powerful  methods to deal with strong interactions. 
Bosonization \cite{tsvelikb} and Density Matrix Renormalization
Group\cite{white} have clarified much of the physics
of the single channel $S=1/2$ Kondo lattice. One finds a 
paramagnetic metallic
phase for small Kondo
coupling and away from half filling. However, conventional 
superconducting fluctuations are strongly
reduced\cite{shibata_dmrg_kondo_1d}. Adding a direct Heisenberg 
interaction
between the spins has been shown to cause an enhancement of
odd-frequency pairing correlations. However, the formation of a spin
gap precluded the observation of fluctuations towards magnetic
ordering\cite{zachar_exotic_kondo}.  

The aim of this paper is to discuss a one  dimensional
Kondo Lattice model with
{\it underscreened moments} in which magnetism can be expected to 
dominate. We
will show that in this one dimensional problem strong fluctuations to
composite superconductivity coexist with strong fluctuations towards
magnetic ordering.

 The generalized Kondo lattice Hamiltonian in one
dimension is:
\begin{eqnarray}
H&=&-t \sum_{i,\sigma} (c^\dagger_{i,\sigma}c_{i+1,\sigma} +
c^\dagger_{i+1,\sigma}c_{i,\sigma})  \nonumber \\ &+&  \lambda_K \sum_i
\vec{S}_i. c^\dagger_{i,\alpha} \vec{\sigma}_{\alpha,\beta}  c_{i,\beta}
+ \sum_i f(\vec{S}_i \cdot \vec{S}_{i+1})
\end{eqnarray}
where $c_{i,\sigma}$ annihilates an electron, $\vec{S}_i$ is a spin-$S$ 
operator, and the function $f$ describes the spin-spin interaction. In 
the
following, we will consider the case $\lambda_K \ll f,t$.

We shall be interested in magnetism and underscreening coexisting with 
superconductivity. To that purpose choose $S \ge 1$.  Several forms of 
magnetic interaction need to be considered in this case.
 The simplest possibility is   to take a Heisenberg
coupling $f(\vec{S}_i \cdot \vec{S}_{i+1})=\lambda_H
\vec{S}_i \cdot\vec{S}_{i+1}$, but no interesting magnetic effects 
would ensue.
  We would
 get at $\lambda_K=0$ either a spin gap for integer $S$ or an
effective spin 1/2 chain for half-odd integer $S$\cite{haldane_gap}.  
Turning on
$\lambda_K\ll t,\lambda_H$, the case of half-odd integer $S$ would 
reduce
to  exact screening, and  a metal with a spin
gap will form\cite{sikkema_spingap_kondo_1d}. The case of integer $S$ 
would lead to a completely trivial
result, the local moments being already completely screened by the
formation of the Haldane gap preventing the Kondo effect and 
leaving the electrons 
essentially free.

Magnetic  effects, on the other hand, would result
from  interactions in the chain, $f$, which lead to low energy dynamics 
described  by effective Hamiltonians richer than the  $SU(2)_{1}$
Wess--Zumino--Novikov--Witten (WZNW) model
which govern the half-odd integer $S$
Heisenberg\cite{tsvelikb} discussed above. We shall discuss systems 
with fixed points
described 
 by  $SU(2)_{2S}$ ($S \ge 1$).
An example is provided by
 the integrable spin-S
chains 
\cite{takhtajan_spin_s,babujian_spin_s,affleck_strongcoupl,affleck_log_c
orr},
where $f =P_S=\sum_{j=1}^S a_j P_j$, $P_j$ is the spin-$j$ projector and
$a_j=\sum_{i=1}^j i^{-1}$. Such $SU_{2S}(2)$  models arise naturally 
under some 
circumstances: it is known 
in particular that the Heisenberg spin-1 chain 
can be described as a $SU(2)_2$ WZNW model perturbed by a mass
term\cite{tsvelik_field} of the order of the Haldane gap. The
generalization of this result to arbitrary spin $S$ chains, based on
a perturbed $SU(2)_{2S}$ WZNW model, was obtained in Ref. 
\onlinecite{cabra_spin_s},
(and will be further discussed below). Thus, our study
should  describe the
Underscreened Kondo--Heisenberg lattice when the Kondo coupling 
exceeds the Haldane gap. For $S=1$,
this corresponds to $0.4\lambda_H < \lambda_K <\lambda_H$.
We shall study the model away 
from half filling and for all $S$, revealing the appearance of a 
critical
 point describing a non Fermi-liquid where magnetism and 
superconductivity
interplay \cite{note_lehur}.

We now proceed to determine the low energy
behavior of the theory. Consider the Hamiltonian in the continuum limit.
According to the
standard prescriptions of non-abelian
bosonization\cite{tsvelikb,witten_wz,affleck_houches}, the electrons are
described by the following  Hamiltonian:
\begin{eqnarray}
H_{\text{el}}&=&H_\rho + H_\sigma \\
H_\rho&=&\int dx \frac{v_F}{2\pi}\left[ (\pi \Pi_\rho)^2 + (\partial_x
\phi_\rho)^2 \right] \nonumber \\
H_\sigma&=&\frac{2\pi v_F}{3} \int dx (\vec{J}_R.\vec{J}_R + 
\vec{J}_L.\vec{J}_L)
\end{eqnarray}
where the canonically conjugate fields $\Pi_\rho,\phi_\rho$ describe
the charge excitations, and the non abelian $SU(2)_1$ currents
$\vec{J}_{R,L}$
describe the electron spin excitations. 
The local moments are described in the low energy regime 
by the Hamiltonian:
\begin{equation}
H_{\text{mom}}=\frac{2\pi v_s}{2(1+S)} \int dx (\vec{S}_R.\vec{S}_R + 
\vec{S}_L.\vec{S}_L)
\end{equation}
where $\vec{S}_{R,L}$ are $SU(2)_{2S}$ WZNW currents.
The Kondo interaction $\lambda_K$  at
incommensurate filling becomes:
\begin{equation}
\lambda_K a \int dx (\vec{J}_R + \vec{J}_L).(\vec{S}_L+\vec{S}_R)(x)
\end{equation}
and preserves spin-charge separation. 
Carrying out standard RG calculations we find that
this interaction is a combination of terms that are
(marginally) relevant in the RG sense and purely marginal terms.
 The former drive
us to a strong coupling fixed point and we  need a
non-perturbative way to determine it. To do so we may
  neglect the
marginal couplings $\vec{J}_R.\vec{S}_R$ and $ \vec{J}_L.\vec{S}_L$
as well as the velocity difference between electron and local moments
spin excitations, and check, subsequently, their relevance at the fixed 
point.
 This  leads us to the following spin Hamiltonian:
\begin{eqnarray}
\label{eq:chiral_hamiltonian}
H_{\text{spin}}&=&H_1+H_2  \\
H_1&=&\int dx \left[ \frac{2\pi v}{2(1+S)} \vec{S}_R.\vec{S}_R + \frac{2
\pi v}{3} \vec{J}_L.\vec{J}_L + \lambda_K a
\vec{S}_R.\vec{J}_L\right]\nonumber \\ 
H_2&=&\int dx \left[ \frac{2\pi v}{2(1+S)} \vec{S}_L.\vec{S}_L + \frac{2
\pi v}{3} \vec{J}_R.\vec{J}_R + \lambda_K a
\vec{S}_L.\vec{J}_R\right] \nonumber
\end{eqnarray}
A similar Hamiltonian was proposed in Refs. 
\onlinecite{azaria_csl,azaria_composite} to
describe the spin sector of a Hubbard chain coupled to $N$ spin 1/2
chains (chain cylinder model). The $N=2$ fixed
point  was analyzed in detail using an exact
solution at a Toulouse point\cite{azaria_csl}, and the correlation 
functions of the composite
order parameters were obtained\cite{azaria_composite}. 

The fixed point Hamiltonian can be determined by arguments
of {\it chiral
stabilization}\cite{andrei_chiral_nfl}, since $H_1$ and $H_2$ in Eq. 
(\ref{eq:chiral_hamiltonian}) are both chirally asymmetric (only their 
sum is chirally
symmetric)\cite{note_andrei}.  We find, that
the electron spin degrees of
freedom are described by the coset CFT - $\frac{SU(2)_1\otimes 
SU(2)_{2S-1}}{ SU(2)_{2S}}$, or equivalently, the minimal
model- $M_{2S-1}$,
of central charge $c=1-\frac{6}{(2S+1)(2S+2)}$, while the local
moments spin
excitations are described by the $SU(2)_{2S-1}$ WZNW model:
\begin{equation}
H^*=\frac{SU(2)_1\otimes SU(2)_{2S-1}}{ SU(2)_{2S}}+SU(2)_{2S-1}.
\end{equation}
This fixed point describes an interesting interplay of magnetism and 
superconductivity. We shall discuss the structure of the ground state, 
the
thermodynamics and then the correlation functions of the model.

The ground state is a coset singlet, formed between electrons and local
moments. A fraction
$\frac{6}{(2S+1)(2S+2)}$ of the electron spin degrees of freedom 
is  absorbed by the spins, leading to its complete screening. This
 manifests itself in the
 magnetic susceptibility and the specific heat:
\begin{eqnarray}
\label{eq:susceptibility}
\chi&=&\frac  1 {2\pi v} (2S-1) \nonumber \\
C&=&\frac{\pi}{3} \left(  1+ \frac{3(2S)}{2S+2}+
1-\frac{12}{(2S+1)(2S+2)}\right) T  \nonumber
\end{eqnarray}
where in the latter quantity we included also charge contributions.
This leads to a Wilson ratio: $
R_W=\frac{10S^2+9S-4}{(2S+1)(S+1)(2S-1)}$
 going to zero as $S\to \infty$.

We
now  calculated the long distance behavior of the physical
correlation functions. To do so we
have  to  express the physical operators in terms
 of the operators around the fixed point, which we proceed
to discuss. The  primary operators of $SU(2)_N$ model are
 $\Phi_N^{(j)}$ ($j\le \frac N 2$) and carry  spin-$j$. The coset
primaries, $\phi^{j
j'}_{j^{\prime \prime}},\;(0\le j
\le 1/2$, $0\le j'
\le S-1/2$ and $0\le  j^{\prime \prime} \le S$), are related to the  
$SU(2)_N$ WZNW spin-$j$ 
primaries via the decomposition:
\begin{equation}
\Phi_{L,1}^{(j)} \Phi_{L,2S-1}^{(j')}=\sum_{|j-j'| \le j^{\prime 
\prime}\le j+j'} \phi^{j,j'}_{L,j^{\prime \prime}}  
\Phi_{L,2S}^{(j^{\prime \prime})}
\end{equation}
 The conformal weight of the coset primary
 is such that the two sides of the equality have
the same conformal weight, implying
 that $\phi^{j
j'}_{L,j^{\prime \prime}}$ has left conformal weight 
$\left(\frac{j(j+1)} 3
 +\frac{j'(j'+1)} {2S+1}- \frac{j^{\prime \prime}(j^{\prime \prime}
 +1)}{2S+2}\right)$. Similar identities hold for the right component.  
Next, note that
at the fixed point products of left primary operators of $SU(2)_1$ and
right $SU(2)_{2S}$ WZNW models have to decompose into a sum of products 
of primary
operators of the left minimal model and  right $SU(2)_{2S-1}$ 
model. Moreover, the total spin has to be conserved. 
Therefore, one can write the decomposition:
\begin{equation}\label{decomposition_ir}
\Phi_{L,1}^{(j)} \Phi_{R,2S}^{(j^{\prime \prime})} \sim 
\sum_{|j-j^{\prime
\prime}| \le j^{\prime}\le j+j^{\prime \prime}}  \phi^{j
j'}_{L, j^{\prime \prime}}   \Phi_{R,2S-1}^{(j')}
\end{equation}
This decomposition satisfies the requirements on the indices in  
$\phi^{j
j'}_{j^{\prime \prime}}$. It is formally similar to a Clebsch-Gordan
decomposition.

Thus, at  the fixed point, the spin operator is given by:
\begin{equation}
\vec{S}_n \sim a (\vec{S}^\prime_R +\vec{S}^\prime_L) (na) +
\phi^{0,1/2}_{1/2} \mathrm{Tr}(\vec{\sigma}g')(na)
\end{equation}
where $\vec{S}^\prime_{R,L}$ and  $g'$ are respectively the currents
and the $SU(2)$ matrix field of the $SU(2)_{2S-1}$
WZNW model, and  $ \phi^{0,1/2}_{1/2}$ is the field of the minimal
model with scaling dimension $\frac{3}{2(2S+1)(2S+2)}$. 
As a result, the spin-spin correlation function behaves as:
\begin{equation}
\langle \vec{S}_n.\vec{S}_m\rangle \sim \frac 1
{(n-m)^2} + \frac{(-)^{n-m}}{(n-m)^{\delta_S}}
\end{equation}
where $\delta_S=\frac {3 (2S+3)}{(2S+1)(2S+2)} \ll 2$. There is  a
strong tendency to antiferromagnetism.

The fermion operator $\psi_L$, is given by:
\begin{equation}
\psi_L=e^{\imath (\theta_\rho+\phi_\rho)/\sqrt2} \phi^{1/2,1/2}_{L \;0} 
g^\prime_R
\end{equation}
where $\theta_\rho(x)=\pi \int^x \Pi_\rho(x') dx'$, 
$\phi^{1/2,1/2}_{L \;0}$ is the
antiholomorphic component of the $\phi^{1/2,1/2}_{0}$ field of the
minimal model, and $ g^\prime_R$ is the holomorphic component of the
$g'$ field of the $SU(2)_{2S-1}$
WZNW model. A similar expression holds for $\psi_R$, in which $L$ and
$R$ are exchanged and $\phi_\rho \to -\phi_\rho$. 
The resulting fermion Green's function is:
\begin{eqnarray}
\langle T c_n(t) c_0^\dagger(0) \rangle \sim \frac {e^{\imath k_F n
a}}{(na-vt)^{1+\frac{3}{2(2S+1)}} (na+vt)^{\frac{3}{2(2S+1)}}} 
\nonumber \\
+ \frac {e^{-\imath k_F n
a}}{(na+vt)^{1+\frac{3}{2(2S+1)}} (na-vt)^{\frac{3}{2(2S+1)}}}
\end{eqnarray}

Considering now  charge density wave (CDW), spin density
wave (SDW),
singlet superconductor (SS) and triplet superconductor (TS)
 correlations, it is easy to see that they  all
decay with the same  exponent $2+\frac{6}{2S+1}$. 
These correlations are thus very strongly suppressed with respect to a
free electron gas. Only for $S\to \infty$, do we 
recover the exponents of the free electron gas for the conventional 
order
parameters. If we specialize to
$N=2$, we obtain the same exponents as in Ref. \onlinecite{azaria_csl}. 

This leads us to investigate the presence of composite
order\cite{coleman_kondo_odd_pair,zachar_exotic_kondo}: the
 composite Charge Density Wave (c-CDW) order parameter $
O_{c-CDW}=\vec{n}(x).\vec{O}_{SDW}(x)$ and the composite singlet order
parameter $
O_{c-S}=\vec{n}(x).\vec{O}_{TS}(x)$.

It is easy to show that the composite Charge Density Wave order
parameter can be expressed at the fixed point as:
\begin{equation}
O_{c-CDW} \sim e^{\imath \sqrt{2} \phi_\rho} \phi^{1/2,0}_{1/2} 
\end{equation}
where the field $\phi^{1/2,0}_{1/2} $ has conformal weight $(\frac 1 4
- \frac 3 {4(2S+2)},\frac 1 4 - \frac 3 {4(2S+2)})$. Similarly,
for the composite singlet order, one has:
\begin{equation}
O_{c-S} \sim e^{\imath \sqrt{2} \theta_\rho} \phi^{1/2,0}_{1/2}. 
\end{equation}
and both correlations decay with the same exponent,
\begin{equation}
\langle O_{composite}(n)O_{composite}^\dagger(n') \rangle \sim \frac 1
{|n-n'|^{2-\frac{3}{2S+2}}} 
\end{equation}
Therefore, for any $S$, composite order parameters are dominant. This
 can be understood in a simple way: electrons are tightly bound to
 local moments, so that composite correlations, that involve  coherent
 motion of local moments excitations and electrons, decay more slowly
 than conventional ones.
 For $S\to \infty$, conventional and composite
order parameters become degenerate. This can be understood by
 remarking that in this limit, the local moments become classical and
 acquire a  non-zero average so that composite and conventional
 order parameters become identical. 

All our discussion up to now was concerned with the coupling of an
integrable spin $S$ chain with fermions. We now turn to  the
effect of the perturbation $\lambda \Phi_R^{(1)}\Phi_L^{(1)}$ that 
restores the behavior of the Heisenberg
spin $S$ chain \cite{cabra_spin_s}. If we assume that this
term is small, i. e. that the Kondo coupling is larger than the
largest gap induced by the perturbation for zero Kondo coupling,
 we can analyze its effect by determining whether it
is relevant or irrelevant at the fixed point. 
The case of the spin 1 chain is special since Kac-Moody selection
rules prevent the appearance of a primary operator of spin 1 in the
fixed point theory. Generalizing (\ref{decomposition_ir}) to include
also non primary operators,  a simple
calculation shows that $\Phi^{(1)}_R \to J_R \xi_L'$, where $J_R$ is
the 
$SU(2)_1$ current, and $\xi_L'$ is a Majorana fermion of the Ising
($M_1$ minimal) model. This result can also be obtained from the 
Toulouse
limit solution\cite{azaria_csl,lecheminant_pc}. 
As a result, the mass term becomes a term of dimension $3$ at
the fixed point and is  irrelevant. We therefore expect that the
underscreened regime we have found will be present if $\lambda_K$ 
exceeds
the Haldane gap of the isolated spin 1 chain. For a spin $S>1$, the 
operator
 $\Phi^{(1)}_{R,2S}$ becomes at the fixed point
$\phi_{L,1}^{0\;1}\Phi^{(1)}_{R,2S-1}$. This operator has dimension
$\frac{8}{2S+1}-\frac 4 {2S+2}<2$ and is therefore \emph{relevant} at 
the
fixed point. It is then likely that for $S>1$
 the models that we have described
flow when perturbed  to the trivial Kondo-Heisenberg fixed point even 
for a
strong Kondo coupling. 

We may also examine what happens when two Kondo chains are coupled 
together to form
a {\it Kondo ladder}. There are several ways to do it. One can 
couple the local moments of the Kondo chains  by a Heisenberg coupling 
$ \lambda_\perp \sum_i
\vec{S}_{i,1}.\vec{S}_{i,2}$. At the fixed point, this gives rise to
relevant terms of 
scaling dimension $\frac{3 (2S+3)}{(2S+1)(2S+2)}$, that induce a spin
gap. Second, one could consider an exchange interaction between the
electrons,  $\lambda_{\text{ee}} \sum_i  c^\dagger_{i,\alpha,1} 
\vec{\sigma}_{\alpha,\beta} 
c_{i,\beta,1}. c^\dagger_{i,\gamma,2} \vec{\sigma}_{\gamma,\delta}
c_{i,\delta,2}$. It is easily seen that such an interaction only leads
to irrelevant terms, and does not affect our fixed point. Finally, one
could consider interchain hopping of the electrons  $t_\perp \sum_i
 (c^\dagger_{i,\sigma,1} c_{i,\sigma,2} +\text{h. c.})$. At the fixed
point, this leads to an operator of dimension
 $1+ \frac{3}{2S+1}$, relevant for $S\ge 3/2$. 
For $S=1$, this term is marginal but it generates a relevant
RKKY interaction between the local moments that destabilizes 
 the chiral fixed point.

We have obtained a physical picture of the one-dimensional
underscreened Kondo Lattice. The formation of a chiral non-Fermi
Liquid results in strong antiferromagnetic fluctuations accompanied
with composite pairing. This picture is reminiscent of the situation
that obtains in 
three dimensional heavy fermion systems\cite{amato_heavy_musr}. It
would be very worthwhile 
to try to determine if some analog of a chiral non-Fermi liquid can be
found in higher dimensional Kondo Lattice models. A good starting
point would be an array of Kondo-underscreened chains coupled by
interchain hopping.

We thank  P. Azaria, P. Lecheminant, H.-Y. Kee, O. Parcollet and A. 
Rosch for
illuminating discussions. E. O. acknowledges support from NSF under
grant DMR 96-14999.
\bibliographystyle{prsty}

\end{document}